\documentclass[dvips]{article}
\newcommand{\doublespacing}{\let\CS=\@currsize\renewcommand{\baselinesstrech}
{2.0}\tiny\CS}
\begin{document}

\textwidth 16cm
\newcommand{\bd}{\begin{document}}
\newcommand{\ed}{\end{document}}
\newcommand{\bc}{\begin{center}}
\newcommand{\ec}{\end{center}}
\newcommand{\bfr}{\begin{flushright}}
\newcommand{\efr}{\end{flushright}}
\newcommand{\lt}{\left}
\newcommand{\rt}{\right}
\newcommand{\vs}{\vspace}
\newcommand{\hs}{\hspace}
\newcommand{\beq}{\begin{equation}}
\newcommand{\eeq}{\end{equation}}
\newcommand{\lb}{\linebreak}
\newcommand{\pb}{\pagebreak}
\newcommand{\mb}{\makebox}
\newcommand{\fb}{\framebox}
\newcommand{\mc}{\multicolumn}
\newcommand{\ben}{\begin{enumerate}}
\newcommand{\een}{\end{enumerate}}
\newcommand{\bit}{\begin{itemize}}
\newcommand{\eit}{\end{itemize}}
\newcommand{\ol}{\overline}
\newcommand{\un}{\underline}
\newcommand{\lefq}{\lefteqn}
\newcommand{\ba}{\begin{array}}
\newcommand{\ea}{\end{array}}
\newcommand{\beqa}{\begin{eqnarray}}
\newcommand{\eeqa}{\end{eqnarray}}
\newcommand{\beqas}{\begin{eqnarray*}}
\newcommand{\eeqas}{\end{eqnarray*}}
\newcommand{\bfg}{\begin{figure}}
\newcommand{\efg}{\end{figure}}
\newcommand{\bds}{\begin{displaymath}}
\newcommand{\eds}{\end{displaymath}}
\newcommand{\btb}{\begin{tabbing}}
\newcommand{\etb}{\end{tabbing}}
\newcommand{\para}{\parallel}
\newcommand{\pad}{\partial}
\newcommand{\nn}{\nonumber}
\newcommand{\la}{\leftarrow}
\newcommand{\ra}{\rightarrow}
\newcommand{\lgla}{\longleftarrow}
\newcommand{\lgra}{\longrightarrow}
\newcommand{\La}{\Leftarrow}\newcommand{\Ra}{\Rightarrow}
\newcommand{\Lra}{\Leftrightarrow}
\newcommand{\Lgla}{\Longleftarrow}
\newcommand{\Lgra}{\Longrightarrow}
\newcommand{\bm}{\boldmath}
\newcommand{\lan}{\langle}
\newcommand{\ran}{\rangle}
\renewcommand{\a}{\alpha}
\renewcommand{\b}{\beta}
\newcommand{\g}{\gamma}
\newcommand{\G}{\Gamma}
\renewcommand{\d}{\delta}
\newcommand{\eps}{\epsilon}
\newcommand{\Th}{\Theta}
\newcommand{\s}{\sigma}
\newcommand{\lam}{\lambda}
\newcommand{\D}{\Delta}
\newcommand{\vare}{\varepsilon}
\newcommand{\pr}{\prime}
\newcommand{\ro}{\rho}
\newcommand{\nab}{\nabla}
\newcommand{\m}{\mu}
\newcommand{\n}{\nu}
\newcommand{\Sg}{\Sigma}
\newcommand{\p}{\pi}
\newcommand{\R}{I\!\!R}
\newcommand{\om}{\omega}
\newcommand{\Om}{\Omega}
\newcommand{\ze}{\zeta}
\newcommand{\vart}{\vartheta}
\newcommand{\tri}{\triangle}
\newcommand{\f}{\frac}
\newcommand{\iny}{\infty}
\newcommand{\pro}{\propto}
\newcommand{\np}{\newpage}
\newcommand {\tab} {\hspace*{2em}}
\newcommand {\lab} {\hspace*{1em}}
\bc {\huge Information Entropy of conditionally exactly solvable potentials} \ec

\vs{1cm}

\bc
{\it D. Dutta{\footnote {e-mail : debjit$_-$r@isical.ac.in} and P. Roy{\footnote{e-mail : pinaki@isical.ac.in}}\\
Physics \& Applied Mathematics Unit \\
Indian Statistical Institute \\
Kolkata - 700 108, India.}} \ec
\vs{4.5cm}

\bc {\large {\un{Abstract}}} \ec
We evaluate Shannon entropy for the position and momentum eigenstates of some conditionally exactly solvable potentials which are isospectral to harmonic oscillator and whose solutions are given in terms of exceptional orthogonal polynomials. The Bialynicki-Birula-Mycielski (BBM) inequality has also been tested for a number of states.
\newpage
\section{Introduction}
In recent years there have been a growing interest in studying information theoretic measures for quantum mechanical systems. In particular entropic uncertainty relations (EUR) which serve as alternatives to Heisenberg uncertainty relation have been examined by a number of authors \cite{dehesa1}.  Among the various measures of information theoretic entropy, the Shannon entropy plays a particularly important role and Shannon entropy of the probability distribution is used as a measure of uncertainty. An EUR relating position and momentum was obtained by Beckner, Bialynicki-Birula and Mycieslki (BBM) \cite{birula} and it is given by
\beq
S_{pos}+S_{mom}\geq D(1+log~\pi)\label{eur}
\eeq
where $D$ denotes the spatial dimension and the position and momentum entropies are defined by
\beq
S_{pos}=-\int |\psi(x)|^2~log|\psi(x)|^2~d^Dx,~~~~S_{mom}=-\int |{\hat\psi}(p)|^2~log|{\hat\psi(p)}|^2~d^Dp\label{Shannon}
\eeq
where $\psi(x)$ is a normalized coordinate space wave function and ${\hat\psi}(p)$ is the corresponding Fourier transform. Apart from their intrinsic interest the EUR (\ref{eur}) or the information entropies (\ref{Shannon}) have been used in different contexts e.g, to study squeezing \cite{or}, localization and fractional revivals \cite{pani,romera} etc. However it is not always easy to exactly evaluate Shannon entropies (\ref{Shannon}) or the EUR (\ref{eur}). They have been obtained exactly only for a few low lying states of the harmonic oscillator \cite{dehesa1,opat}, P\"oschl-Teller \cite{dehesa2,coffey}, Morse potential\cite{dehesa2,sever}, Coulomb potential etc \cite{dehesa3} and for other states especially those with large quantum numbers numerical results have been obtained \cite{dehesa3}. In a related development position space entropy for some potential isospectral to the P\"oschl-Teller potential has also been computed \cite{anil}. Also information entropies of several classical orthogonal polynomials e.g, Laguerre, Hermite, Gegenbauer etc., which provide solutions of most of the standard solvable potentials have been calculated \cite{dehesa4}. However there are many exactly solvable potentials, especially the isospectral partners of the standard ones whose solutions can not be expressed in terms of classical orthogonal polynomials and as far as information theoretic measures are concerned not much is known about these potentials.

Very recently a new class of orthogonal polynomials, called the exceptional orthogonal polynomials \cite{ulate}, have been studied by a number of authors \cite{quesne}. These polynomials are different from the classical orthogonal polynomials and have rather distinct properties. Interestingly exceptional orthogonal polynomials multiplied by square root of the associated weights appear as solutions of a number of exactly solvable potentials. Here we shall examine a class of conditionally exactly solvable potentials \cite{dutra} which are isospectral to the harmonic oscillator potential \cite{roy1,roy2} and whose solutions are given in terms of exceptional orthogonal polynomials related to the generalized Laguerre and Hermite polynomials \cite{roy3}. These potentials consist of a harmonic oscillator term and other terms of a non polynomial type and may be viewed as a deformation of the harmonic oscillator \cite{gonzales}. In view of the fact that the isospectral partner potentials have rather different type of eigenfunctions it is of interest to evaluate their position and momentum space entropies and compare them with the same of their partners. In this paper we shall consider the models of ref \cite{roy3} and examine them from the point of view of EUR (\ref{eur}).

\section{Exceptional orthogonal polynomials associated with Conditionally exactly solvable potentials}
It may be noted that a supersymmetric quantum mechanical system consists of a pair of Hamiltonians $(H^{\pm})$ of the form (in units of $\hbar=m=1$)  \cite{khare}
\beq
H^{\pm}=A^{\pm}A^{\mp}=-\f{1}{2}\f{d^2}{dx^2}+V_{\pm}(x),~~~~A^{\pm}=\f{1}{\sqrt{2}}\left(\pm\f{d}{dx}+W(x)\right),~~~~V_{\pm}(x)=\f{1}{2}\left(W^2(x)\pm W^\prime(x)\right)
\eeq
Thus zero energy states of the Hamiltonians $H_\pm$ can be obtained from the solutions of the equations $A^\mp \psi_0^\pm=0$ and they are given by $\psi_0^\pm=e^{\pm\int W(x)dx}$. If either of $\psi_0^\pm$ is normalizable supersymmetry is unbroken while if neither is normalizable then supersymmetry is said to be broken. However both of $\psi_0^\pm$ can not be normalizable at the same time (at least in the examples we shall consider). So if supersymmetry is unbroken then the above Hamiltonians are isospectral except the zero energy state (which is assumed to belong to $H_-$). In this case the relationship between the energies and the eigenfunctions of the Hamiltonians $H^{\pm}$ are given by
\beq
E_0^-=0,~~~~E_{n+1}^-=E_n^+>0\label{3}
\eeq
\beq
\psi_0^-=N~e^{-\int W(r)dr},~~~~\psi_n^+ = \f{1}{\sqrt{E_{n+1}^-}}~A^+\psi_{n+1}^-,~~~~\psi_{n+1}^- = \f{1}{\sqrt{E_{n}^+}}~A^-\psi_n^{+}\label{4}
\eeq

Thus all the states are doubly degenerate except the zero energy ground state. On the other hand when supersymmetry is broken we have
\beq
E_n^+=E_n^->0,~~~~\psi_n^+=\f{1}{\sqrt{E_n^-}}~A^+\psi_n^-,~~~~\psi_n^-=\f{1}{\sqrt{E_n^+}}~A^-\psi_n^+\label{broken1}
\eeq
Thus in this case all the states including the ground state are doubly degenerate. We shall now examine the case of broken supersymmetry and consider the following superpotential :
\beq
W(r)=r+\f{l+1}{r}+\f{u^\prime(r)}{u(r)},~~~~0<r<\infty,~~~~l\geq 0
\eeq
where $u(r)$ is given by
\beq
u(r)={}_1{F}_1\left(\f{1-\eps}{2},l+\f{3}{2},-r^2\right)
\eeq
It is seen that $u(r)$ will be a polynomial for any positive odd integral value of $\eps$. It can be checked that neither of $\psi_0^{\pm}$ is normalizable and consequently supersymmetry is broken. Now we choose the simplest non trivial value i.e, $\eps=3$ so that
\beq
u(r)= \f{2r^2+2l+3}{2l+3}
\eeq
Then it can be shown that \cite{roy2,roy3}
\beq
V_+(r)=\f{r^2}{2}+\f{l(l+1)}{2r^2}+l+\f{7}{2}\label{vp}
\eeq
\beq E_n^+=E_n^-=2n+2l+5,~~~~\psi_{n}^{+}=\sqrt{\f{2(n!)}{\G(n+l+\f{3}{2})}}~r^{l+1}L_n^{l+\f{1}{2}}(r^2)e^{-r^2/2},~~n=0,1,\cdots
\eeq

\beq
V_-(r)=\f{r^2}{2}+\f{(l+1)(l+2)}{2r^2}+\f{4r}{(2r^2+2l+3)}\left(2r+\f{2(l+1)}{r}+\f{4r}{2r^2+2l+3}\right)+l-\f{3}{2}\label{vm}
\eeq
\beq
\psi_{n}^{-}=\sqrt{\f{n!}{(2n+2l+5)\G(n+l+\f{3}{2})}}~\f{e^{-\f{r^2}{2}}r^{l+2}}{(2r^2+2l+3)}\left[ 4L_n^{l+\f{1}{2}}(r^2)+2(2r^2+2l+3)L_n^{l+\f{3}{2}}(r^2)\right]\label{wf-}
\eeq
There are several points to be observed. First to be noted is that the solutions (\ref{wf-}) can be expressed in terms of the following exceptional orthogonal polynomials and the weight function in the following way \cite{roy3}:
\beq
\ba{l}
\psi_n^-(r)=N_n\sqrt{w(r)}~p_n(r)\\
p_n(r)=\displaystyle\f{1}{(2l+3)}\left[ 4L_n^{l+\f{1}{2}}(r^2)+2(2r^2+2l+3)L_n^{l+\f{3}{2}}(r^2)\right]\\
w(r)=\displaystyle\left(\f{2l+3}{2r^2+2l+3}\right)^2e^{-r^2}r^{2l+4}\ea\eeq
Secondly $V_+(r)$ is the effective potential for the radial oscillator while $V_-(r)$ is the conditionally exactly solvable \footnote[1]{If a potential is exactly solvable only if some of the couplings in the potential assume specific values, then such a potential is termed as conditionally exactly solvable \cite{dutra}.} isospectral partner which is non shape invariant.

\section{Information entropy for the potential $V_-(r)$}
It may be noted that so far we have considered a system on the half line and $l$ is just a parameter. However if we wish to view our system as a three dimensional system with a radially symmetric potential then $l$ has to be treated as angular momentum. It may be observed that the radial part of $V_+(r)$ is that of a standard radial oscillator while that of $V_-(r)$ actually depends on $l$ and consequently the degeneracy pattern for a fixed value of $l=L$ is given by $E^+_{n,L+1}=E^-_{n+1,L}$. Nevertheless it will be seen that the operators $A^+$ or $A^-$ may be used to simplify calculation of the position entropy of the $(-)$ sector.  We would like to note that $V_+(r)$ represents the effective potential for the radial harmonic oscillator. On the other hand the potential $V_-(r)$ is isospectral partner of $V_+(r)$. However $V_-(r)$ is of anharmonic type and it is not shape invariant. Here our objective is to evaluate Shannon entropy for the isospectral partner $V_-(r)$ and examine their differences with those of $V_+(r)$.

The potential $V_-(r)$ in (\ref{vm}) and the wave function (\ref{wf-}) are the effective potential and the reduced wave function. The actual spherically symmetric potential is the same as (\ref{vm}) except the angular momentum term while the complete wave functions are given by
\beq
\psi_{n,l+1,m}^{-}=\sqrt{\f{n!}{(2n+2l+5)\G(n+l+\f{3}{2})}}~\f{e^{-\f{r^2}{2}}r^{l+1}}{(2r^2+2l+3)}\left[ 4L_n^{l+\f{1}{2}}(r^2)+2(2r^2+2l+3)L_n^{l+\f{3}{2}}(r^2)\right]~Y_{l+1}^m(\theta,\phi)\label{cwf-}
\eeq
where $Y_{l+1}^m(\theta,\phi)$ denote the angular part of the wave functions. The position entropy can be readily evaluated using (\ref{cwf-}). To evaluate the momentum entropy it is necessary to obtain the Fourier transform of (\ref{cwf-}). In this context it may be noted that momentum space wave functions for the harmonic oscillator or the Coulomb potential can be obtained in a relatively simple manner \cite{flugge}. However for other potentials, particularly the isospectral partners of the standard potentials, the momentum space wave functions can not be obtained in the same way. Here our approach would be to express the momentum space wave functions of $V_-(r)$ partly in terms of the momentum space wave functions of the radial oscillator which are already known \cite{dehesa3,flugge}. To this end we note that the Fourier transform of the wave function (\ref{cwf-}) is given by \cite{flugge}
\beq
{\hat \psi}_{n,l+1,m}^-(p)=\f{1}{p}g_{l+1}(p)Y_{l+1}^m(\Theta,\Phi)
\eeq
where
\beq
g_{l+1}(p)=\sqrt{\f{2}{\pi}}~i^{-(l+1)}\int_0^\infty j_{l+1}(pr)\psi_{n,l}^-~dr,~~~~j_l(z)=\sqrt{\f{\pi z}{2}}J_{l+\f{1}{2}}(z),\label{gl}
\eeq
$J_{l+\f{1}{2}}(z)$ being the Bessel function (of the first kind) of order $(l+\f{1}{2})$ \cite{grads}. Now using the intertwining relation (\ref{broken1}) inside the integral (\ref{gl}) we obtain
\beq
{\hat\psi}_{n,l+1,m}^-(p)=\sqrt{\f{1}{2n+2l+5}}\left[\sqrt{2n+2l+3}~{\hat\psi}_{n,l+1,m}^+(p)+\f{i^{-(l+1)}}{p}\sqrt{\f{32n!}{\pi\G(n+l+\f{3}{2})}}~Y_{l+1}^m(\Theta,\Phi)~I(p)\right]\label{fourier1}
\eeq
where ${\hat\psi}_{n,l+1,m}^+(p)$ is the Fourier transform of $\psi_{n,l+1,m}^+(r)$ \cite{dehesa3,flugge} :
\beq
{\hat\psi}_{n,l+1,m}^+(p)=(-i)^{l+1}\sqrt{\f{2n!}{\G(n+l+\f{5}{2})}}~p^{l+1}e^{-p^2/2}L_n^{l+3/2}(p^2)Y_{l+1}^m(\Theta,\Phi)
\eeq
The function $I(p)$ is the solution of the differential equation
\beq
\f{d^2I(p)}{dp^2}-\left[\f{(l+1)(l+2)}{p^2}+(l+\f{3}{2})\right]I(p)+\f{1}{2}(-1)^n\sqrt{\f{\pi}{2}}~p^{l+2}e^{-p^2/2}\left[L_n^{l+\f{3}{2}}(p^2)+L_{n-1}^{l+\f{3}{2}}(p^2)\right]=0
\eeq and is given by
\beq
\ba{l}
I(p)=\displaystyle\sqrt{p}\left[J_{l+3/2}(-i\sqrt{l+3/2}p)\left(C_1+\f{(-1)^n\pi^{3/2}}{4\sqrt{2}}\int_1^pe^{-t^2/2}Y_{l+3/2}(-i\sqrt{l+3/2}t)t^{l+5/2}(L_{n-1}^{l+\f{3}{2}}(t^2)+L_{n}^{l+\f{3}{2}}(t^2))dt\right)\right.\\
\displaystyle\left.+Y_{l+3/2}(-i\sqrt{l+3/2}p)\left(C_2-\f{(-1)^n\pi^{3/2}}{4\sqrt{2}}\int_1^pe^{-t^2/2}J_{l+3/2}(-i\sqrt{l+3/2}t)t^{l+5/2}(L_{n-1}^{l+\f{3}{2}}(t^2)+L_{n}^{l+\f{3}{2}}(t^2))dt\right)\right]\label{ip}
\ea
\eeq
where $Y_l(z)$ denotes Bessel function of the second kind \cite{grads}. The constants $C_{1,2}$ depend on the quantum numbers $n,l$ and have to be determined from the boundary conditions
\beq
\lim_{p\rightarrow 0,\infty} {\hat\psi}_{n,l+1,m}^-(p)=0\label{limit}
\eeq

It may be pointed out that the expression (\ref{fourier1}) for the momentum eigenstates is an exact one. Now using (\ref{cwf-}) and (\ref{fourier1}) we shall compute the Shannon entropies (\ref{Shannon}) and examine the EUR (\ref{eur}) for different values of the quantum number $n$ and the results are given in Tables 1 and 2. We would like to note that in Tables 1 and 2 we have considered those eigenfunctions which correspond to the same energy. From Table 1 it is seen that the BBM inequality is always satisfied as it should be and the sum of entropies increases with $n$. From Table 2 we find that except for $n=0$, for the same energy the sum of the entropies for the isospectral partner (\ref{vm}) is less than the same for the original potential. 
\subsection{Information entropy for isospectral partner of linear harmonic oscillator}
Here we take the superpotential to be \cite{roy1,roy2}
\beq
W(x)=x+\f{4x}{1+2x^2},~~~~-\infty<x<\infty
\eeq
Then the partner potentials and their solutions are given by
\beq
V_+(x)=\f{x^2}{2}+\f{5}{2},~~~~\psi_n^+=\sqrt{\f{1}{2^nn!\sqrt{\pi}}}~e^{-x^2/2}H_n(x),~~~~E_n^+=n+3,~~~~n=0,1\cdots\label{unbrokenx}
\eeq
\beq
\ba{l}
V_-(x)=\displaystyle\f{x^2}{2}-\f{4}{1+2x^2}+\f{16x^2}{(1+2x^2)^2}+\f{3}{2},~~~~\\\\
\displaystyle\psi_0^-=\sqrt{\f{2}{\sqrt{\pi}}}~\f{e^{-x^2/2}}{(1+2x^2)},~~~~E_0^-=0\\\\
\psi_{n+1}^-=\displaystyle \f{1}{\sqrt{2^{n+1}n!(n+3)\sqrt{\pi}}}~\f{e^{-x^2/2}}{(1+2x^2)}\left[(1+2x^2)H_{n+1}(x)+4xH_n(x)\right],~~~~E_{n+1}^-=n+3
\ea\label{unbrokenx1}
\eeq
In this case the polynomials and the weight function are given by
\beq
p_0(x)=1,~~~~p_n(x)=(1+2x^2)H_{n+1}(x)+4xH_n(x),~~~~\omega(x)=\f{e^{-x^2}}{(1+2x^2)^2}
\eeq
It may be noted that $V_+(x)$ in (\ref{unbrokenx1}) is a linear harmonic oscillator (with a shifted energy scale) while its partner is a conditionally exactly solvable potential. A major difference with the previous example is that in the present case $V_-(x)$ has a zero energy ground state and consequently supersymmetry is unbroken. Thus the ground state of $V_-(x)$ is a singlet while the excited states are degenerate. Also because of the relations (\ref{4}) it is clear that the ground state $\psi_0^-$ is annihilated by both $A^+$ and $A^-$ and has to be considered separately.  Following the procedure of the last section it can be shown that
\beq
{\hat \psi}_0^-(p)=\f{(\pi e)^{1/4}}{2\sqrt{2}}\left[2cosh(\f{p}{\sqrt{2}})-e^{-\f{p}{\sqrt{2}}}~erf[\frac{1}{2}-\f{p}{\sqrt{2}}]-e^{{p\sqrt{2}}}~erf[\frac{1}{2}+\f{p}{\sqrt{2}}]\right]\label{unbrokenp}
\eeq
\beq
{\hat \psi}_{n+1}^-(p)=\f{1}{\sqrt{2(n+3)2^n n!\sqrt{\pi}}}\left[\sqrt{\sqrt{\pi} 2^{n+1}(n+1)!}~{\hat\psi}_{n+1}^+(p)+4I(p)\right]\label{unbrokenp1}
\eeq
where 
\beq
{\hat\psi}_n^+(p)=\f{1}{\sqrt{2^nn!\sqrt{\pi}}}e^{-p^2/2}H_n(p)
\eeq
is the momentum space wave functions of the potential (\ref{unbrokenx}) while $I(p)$ is the solution of the differential equation
\beq
\f{d^2I(p)}{dp^2}-\f{1}{2}I(p)+\f{i^{n-1}e^{-p^2/2}}{\sqrt{2^{n+2}n!\sqrt{\pi}}}\left[2nH_{n-1}(p)-pH_n(p)\right]=0
\eeq
and is given by
\beq
\ba{l}
I(p)=\displaystyle e^{-\f{p}{\sqrt{2}}}\left[C_2+e^{\sqrt{2}p}\left(C_1+\f{i^{n+1}}{2\sqrt{2^{n+1}n!\sqrt{\pi}}}\int_1^p e^{-\f{1}{2}q(\sqrt{2}+q)}(2nH_{n-1}(q)-qH_n(q))dq\right)\right.\\
\displaystyle\left.-\f{i^{n+1}}{2\sqrt{2^{n+1}n!\sqrt{\pi}}}\int_1^p e^{\f{1}{2}q(\sqrt{2}-q)}(2nH_{n-1}(q)-qH_n(q))dq\right]
\ea
\eeq 
As in the previous example the constants $C_{1,2}$ depend on the quantum number $n$ and have to be determined for each $n$ from the condition
\beq
\lim_{p\rightarrow\pm\infty} {\hat \psi}_{n+1}^-(p)=0
\eeq
Using (\ref{unbrokenx})-(\ref{unbrokenp1}) we have calculated position and momentum entropies for a few levels and the results are given in Table 3. From Table 3 it is seen that in contrast to $V_-(x)$, the ground state of the isospectral partner does not saturate the BBM inequality. Also for states corresponding to the same energy position and momentum entropies of the isospectral partner is more than the same of the linear oscillator.

\section{Conclusion}
In this paper we have computed the Shannon entropies for the position and momentum eigenstates for a first few levels of two conditionally exactly solvable potentials whose solutions are given in terms of exceptional orthogonal polynomials. It has been found that for linear oscillator the entropies of the eigenstates of the isospectral potential are more than those of the linear oscillator. However in the radial problem the situation is similar for the initial levels although the angular momentum of the eigenfunctions differ by one unit. 

Here we have considered a certain type of potential and we feel it would be of interest to examine other potentials whose solutions are given in terms of exceptional orthogonal polynomials of other types e.g, exceptional Jacobi polynomials \cite{quesne}. It may also be noted that we have obtained the results for the some initial states and it is desirable to be able to evalaute entropies for any value of the quantum number(s). However computation of entropies for large quantum numbers is a formidable job. In view of this we feel it would be interesting to study asymptotic behaviour (with respect to the quantum number) of the entropies of exceptional orthogonal polynomials. Finally we would like to mention that it would be worth investigating some other properties like the Fisher information \cite{fisher}, spreading length \cite{spread} etc. of this new class of polynomials.

\newpage
\begin{center}
\begin{tabular}{|c|c|c|c|c|}

  \hline n & $S^{-}_{pos}(n)$&$ S^{-}_{mom}(n)$ & $S^{-}_{pos}(n)+S^{-}_{mom}(n)$ &$3(1+log(\pi))$\\

 \hline  0 & 3.361 & 3.646 & 6.917&6.434\\
 \hline  1& 4.015 & 4.199 & 8.214& 6.434\\
 \hline  2& 4.568 & 4.628 & 9.196 &6.434\\
 \hline  3 &4.822 & 4.954 & 9.776&6.434\\
  \hline
\end{tabular}
\end{center}
\hspace*{3.1cm}{{\bf{{Table 1. Information Entropies and their sum for the}\\
\hspace*{3.1cm}{eigenstates of $V_-(r)$ in (\ref{vm}) for $l=m=0$.}}}}
\vspace{2.2cm}

\begin{center}
\begin{tabular}{|c|c|c|c|c|}

  \hline n & $S^{+}_{pos}(n)$&$ S^{-}_{pos}(n)$ & $S^{+}_{mom}(n)$& $S^{-}_{mom}(n)$ \\

 \hline  0 & 3.217 & 3.361& 3.217&3.646\\
 \hline  1& 4.151 & 4.015& 4.151& 4.199\\
 \hline  2& 4.709 & 4.568& 4.709 &4.628\\
 \hline  3 & 5.109 & 4.822 & 5.109&4.954\\
  \hline
\end{tabular}
\end{center}
\hspace*{4.4cm}{\bf{Table 2. Information Entropies for the\\
\hspace*{4.4cm}eigenstates of the potentials in (\ref{vp}) and (\ref{vm})\\
\hspace*{4.4cm}for $l=m=0$.
}}
\vspace{2.2cm}
\begin{center}
\begin{tabular}{|c|c|c|c|c|c|c|c|}

  \hline n & $S^{+}_{pos}(n-1)$&$S^{+}_{mom}(n-1)$&$S^{+}_{pos}(n-1)$&$S^{-}_{pos}(n)$&$ S^{-}_{mom}(n)$ & $S^{-}_{pos}(n)$ &$1+log(\pi)$\\
  &&&$+ S^{+}_{mom}(n-1)$&&&$+S^{-}_{mom}(n)$&\\

 \hline  0&-&-&- & 0.479 & 1.679 & 2.158&2.144\\
 \hline  1&1.072&1.072&2.144& 1.261 & 1.607& 2.868& 2.144\\
 \hline  2&1.343&1.343&2.686& 1.425 & 1.578 & 3.003 &2.144\\
 \hline  3 &1.499&1.499&2.998& 1.578 & 1.748 &3.326&2.144\\
  \hline

\end{tabular}
\end{center}

\hspace*{.001cm}{\bf{{Table 3. Information Entropies and their sum for the}
eigenstates of the potentials\\ 
\hspace*{.3cm} in (\ref{unbrokenx}). }}

\ed